\documentclass[superscriptaddress,twocolumn,showpacs,prl,amsmath,amssymb]{revtex4}

\usepackage[varg]{txfonts}

\usepackage{graphicx}
\usepackage{bm}

\def\XXint#1#2#3{{\setbox0=\hbox{$#1{#2#3}{\int}$}
    \vcenter{\hbox{$#2#3$}}\kern-.5\wd0}}

\begin{document}

\title{Non-perturbative microscopic theory of superconducting
fluctuations near a quantum critical point}

\author{Victor~Galitski}
\affiliation{Department of Physics and Joint Quantum Institute,
University of Maryland, College Park, MD 20742-4111}

\begin{abstract}
We consider an inhomogeneous anisotropic gap superconductor in the
vicinity of the quantum critical point, where the transition
temperature is suppressed to zero by disorder. Starting with the
BCS Hamiltonian, we derive the Ginzburg-Landau action for the
superconducting order parameter. It is shown that the critical
theory corresponds to the marginal case in two dimensions and is
formally equivalent to the theory of an antiferromagnetic quantum
critical point, which is a quantum critical theory with the
dynamic critical exponent, $z=2$. This allows us to use a parquet
method to calculate the non-perturbative effect of quantum
superconducting fluctuations on thermodynamic properties. We
derive a general expression for the fluctuation magnetic
susceptibility, which exhibits a crossover from the logarithmic
dependence, $\chi(T,H,n) \propto \ln \left[ \delta n(T,H)\right]$,
valid beyond the Ginzburg region to $\chi(T,H,n) \propto \ln^{1/5}
\left[ \delta n(T,H)\right]$ valid in the immediate vicinity of
the transition (where $\delta n$ is the deviation from the
critical disorder concentration).  We suggest that the obtained
non-perturbative results describe the low-temperature critical
behavior of a variety of diverse superconducting systems, which
include overdoped high-temperature cuprates, disordered $p$-wave
superconductors, and conventional superconducting films with
magnetic impurities.
\end{abstract}

 \maketitle

The problem of the critical behavior of itinerant electronic
systems in the vicinity of quantum phase transitions has been the
subject of intense theoretical investigations in recent
years~\cite{QPT.book,QPT.RMP,HMM,Millis,CPS}. Most theoretical
studies of the problem are based on effective field theories,
which describe critical fluctuations near the transition. The bare
Ginzburg-Landau coefficients of these models are usually treated
as phenomenological parameters, which often remain undefined due
to the lack of a reliable microscopic theory of the transition.
Indeed, while Fermi liquid theory predicts and classifies possible
intrinsic instabilities, there is usually no controlled approach
to access the transition point on the basis of a fermionic
Hamiltonian. The notable exception is a superconducting
instability, which is well-described by the perturbative
Bardeen-Cooper-Schrieffer (BCS) theory. A strong magnetic field or
disorder effects may suppress the superconducting transition
temperature to zero and therefore lead to a superconducting
quantum critical point (QCP), which can be studied on the basis of
the BCS model.

The critical behavior of a superconducting system near the
transition is governed by superconducting fluctuation
effects~\cite{LV}, which physically are due to uncondensed
fluctuating Cooper pairs, which co-exist with electronic
excitations. A perturbative theory of classical superconducting
fluctuations was developed  by Aslamazov and Larkin in
1968~\cite{AL}. More recently, Larkin and the author~\cite{GaL1}
considered quantum superconducting fluctuations near the
magnetic-field-tuned QCP. In both cases, it was found that
Gaussian fluctuations strongly affect thermodynamics and transport
near the critical point.  Even though the Alsamazov-Larkin theory
has been very successful in describing a variety of experiments,
it is strictly speaking not a critical theory, but a Gaussian
perturbation theory, which assumes that the fluctuating Cooper
pairs do not interact, and applies only far enough from the
transition (beyond the Ginzburg region). To the best of the
author's knowledge, there is no physically relevant example of a
non-perturbative treatment of superconducting fluctuations.

In this paper, we develop such a non-petrubative microscopic
theory for a disordered anisotropic gap superconductor near the
disorder-tuned QCP. We derive the corresponding critical theory,
which is shown to be identical to the Hertz-Moriya-Millis
theory~\cite{HMM,Millis} of an antiferromagnetic QCP in two
dimensions. We find that the dimensionless bare quartic coupling
(which characterizes the interaction between superconducting
fluctuations) is a small parameter of the order of the inverse
conductance  and  this quartic term becomes marginally
irrelevant at the transition. 
This allows us to perform non-perturbative parquet resummation of
the leading logarithms and find the exact critical behavior of the
magnetic susceptibility near the transition. The latter crosses
over from  $\chi(T,n) \propto \ln \left[ n - n_{\rm
c}(T)/n\right]$ (which is the quantum analogue of the
Aslamazov-Larkin result valid far enough from the transition) to
the critical behavior $\chi(T,n) \propto \ln^{1/5} \left[ n -
n_{\rm c}(T)/n\right]$, which holds
within the quantum Ginzburg region. 

Let us consider a disordered electron system with a weak
attraction in the $l$-wave channel, described by the Hamiltonian
\begin{eqnarray}
\label{H}
 \hat{\cal H} = \sum\limits_{\bf p}  \hat{\psi}^{\dagger}_{\bf p}
\xi_{\bf p} \hat{\psi}_{\bf p} + {1 \over 2} \sum\limits_{{\bf p},
{\bf p}', {\bf q}} V({\bf p},{\bf p}') \hat{\psi}^{\dagger}_{\bf
p} \hat{\psi}^{\dagger}_{-{\bf p} + {\bf q}} \hat{\psi}_{{\bf p}'}
\hat{\psi}_{-{\bf p}' + {\bf q}} +{\hat{\cal H}}_{\rm dis},
\end{eqnarray}
where $\xi_{\bf p} = E({\bf p}) - \mu$ is the spectrum,
${\hat{\cal H}}_{\rm dis}$ represents a disorder potential (which
we assume to be due to Poisson distributed short-range
impurities), the interaction is $V({\bf p},{\bf p}') = - \lambda_l
u_l(\phi) u_l(\phi)$, $\lambda_l$ is the interaction constant, and
$u_l(\phi)$ defines the symmetry of the gap. In what follows we
assume an unusual pairing symmetry ({\em  e.g.}, $d$-wave), so
that $\overline{u_l} = \int_0^{2\pi} u_l(\phi) d\phi/(2 \pi) = 0$
and $\overline{u_l^2} = 1$. We suppress spin indices throughout
the paper. Next, we introduce an anisotropic order parameter and
allow for its spatial and temporal fluctuations
\begin{equation}
\label{D.def} \Delta_{\bf k} ({\bf r},\tau) = T \sum\limits_{{\bf
k}', {\bf q}, \omega_n} V({\bf k},{\bf k}') F\left({\bf k}' -
{{\bf q} \over 2},{\bf k} + {{\bf q} \over 2}; \omega_n\right)
e^{i {\bf q} \cdot {\bf r} -i \omega_n \tau },
\end{equation}
where $\tau$ is the Matsubara time and $F({\bf r},\tau; {\bf
r}',\tau') = -\left\langle T_{\tau} \hat{\psi} ({\bf r},\tau)
\hat{\psi} ({\bf r},\tau) \right\rangle$ is the Gor'kov's Green's
function.  Following Ref.~[\onlinecite{VG}], we assume that the
symmetry of the gap is preserved, $\Delta_{\bf k} ({\bf r},\tau) =
\Delta ({\bf r},\tau) u_l(\phi_{\bf k})$, and   integrate out the
one-particle degrees of freedom to obtain the following effective
action for the order parameter near the transition
\begin{eqnarray}
\label{S.eff} S[\Delta] =&& \nu \int_Q \left[ {\tau_c (T,H) - \tau
\over \tau_{\rm c0} }  + |\omega| \tau + D
{\bf q}^2 \tau \right] \left| \Delta (Q) \right|^2 \\
\nonumber && \!\!\!\!\!\!\!\!\!\! + {B \over 2} \int_{Q_1,Q_2,Q_3}
\Delta^* (Q_1) \Delta (Q_2) \Delta^* (Q_3) \Delta (Q_1 +Q_3 -
Q_2),
\end{eqnarray}
where $\nu$ is the density of states at the Fermi line and we use
a three-vector $Q = (\omega_n,{\bf q})$ and the notation $\int_Q
f(Q) \equiv T \sum\limits_{\omega_n} \int {d^2{\bf q}/(2 \pi)^2}
f(\omega_n,{\bf q})$ for brevity ($\omega_n$ is the bosonic
Matsubara frequency). The general expression for the quartic
coefficient $B$ (at an arbitrary temperature and disorder
strength) was derived by the author in Ref.~[\onlinecite{VG}].
Near the QCP, $T \tau_{\rm c0} \to 0$, it reduces to $B = \nu
\tau^2 ( \overline{u_l^4} - 1/3)$ [{\em c.f.}, $B = {7
\overline{u_l^4} \zeta(3) \nu / (8 \pi^2 T^2)}$ in the classical
limit, $T \tau_{\rm c0} \gg 1$], where the overline implies an
averaging over the directions on the Fermi line. {\em E.g.}, in a
 $d$-wave superconductor, we get $\overline{u_{\rm d}^4} = 3/2$.
In Eq.~(\ref{S.eff}), the critical scattering time  (or disorder
concentration, $n_{\rm c} \sim \tau_{\rm c}^{-1}$) determines the
superconducting transition point and is a function of temperature
and magnetic field. To find the dependence on the latter, we can
just replace the Cooper-pair momentum ${\bf q}$ with the operator
$\hat{\bm \pi} = \left[ - i {\bm \nabla} - {2e \over c} {\bf
A}({\bf r}) \right]$ and evaluate the matrix element corresponding
to the lowest Landau level in Eq.~(\ref{S.eff}) by replacing $D
{\bf q}^2$ with $\left\langle 0 | D {\hat{\bm \pi}}^2 | 0
\right\rangle = 2 e D H/c$. The full three-dimensional
temperature-disorder-magnetic field phase diagram is determined by
the  equation
\begin{equation}
\label{AG} \ln{T_{c0} \over T} = \psi \left\{ {1 \over 2} + {1
\over 4 \pi T} \left[ {1\over \tau_c(T,H)} + 2{e \over c} DH
\right] \right\} - \psi \left( {1 \over 2} \right),
\end{equation}
where $\psi(z)$ is the logarithmic derivative of the
gamma-function. The zero-temperature solution of this equation
represents a line of quantum critical points. In what follows, we
 consider the low-temperature limit and
 small magnetic fields. In this case,
the asymptotic behavior of the critical scattering time is
\begin{equation}
\label{QCP} { \tau_{\rm c}(T,H) - \tau_{\rm c0}  \over \tau_{\rm
c0}} = 2 \pi g \omega_{\rm c} \tau_{\rm c0} + {2 \gamma^2 \over 3}
\left( {T \over T_{\rm c0}}\right)^2,
\end{equation}
where $g = E_{\rm F} \tau /\pi \gg 1$ is the dimensionless
conductance, $\omega_{\rm c} = {e H / (mc)}$ is the cyclotron
frequency, $1/\tau_{\rm c0} = \pi T_{\rm c0}/\gamma$ is the
critical scattering rate corresponding to $T=0$ and $H = 0$,
$\gamma \approx 1.781$ is the Euler's constant, and $T_{\rm c0}
\propto \exp[-1/(\lambda_l \nu)]$ is the BCS transition
temperature in a clean system.

 Now, we consider the ultra-low-temperature
regime in which quantum rather than thermal fluctuations determine
the critical behavior of the system, which corresponds to the
limit $r(T,H,\tau) = {\left[ \tau_{\rm c}(T,H) - \tau\right] /
\tau} \gg 2 \pi T \tau$. In this case, we can replace the
Matsubara sums in Eq.~(\ref{S.eff}) with the integrals over the
frequency. It is also convenient to introduce new variables $r =
{\left[ \tau - \tau_{\rm c}(T,H) \right]/ \tau}$ and $\rho_0 = B /
(8 D \tau^2 \nu^2)$ and rescale the parameters in
Eq.~(\ref{S.eff}) as follows: $k_0 = \omega \tau$, ${\bf k} =
\sqrt{D \tau} {\bf q}$, and $\phi = \sqrt{2 \nu/(D t^2)} \Delta$.
Using these dimensionless variables, we arrive to the following
quantum critical action
\begin{eqnarray}
\label{S} S[\phi] =&& {1 \over 2!} \int_k \left[ r + |k_0| +
{\bf k}^2  \right] \left| \phi (k) \right|^2 \\
 && \nonumber\!\!\!\!\! + {\rho_0 \over 4!}
\int_{k_1,k_2,k_3} \phi^* (k_1) \phi (k_2) \phi^* (k_3) \phi (k_1
+k_3 - k_2),
\end{eqnarray}
where the parameter $\rho_0 = ( 3 \overline{u_l^4} - 1 )/g \ll 1$
 is essentially the inverse conductance and thus is small.
We emphasize that the above action is not a phenomenological
effective theory, but a microscopic result, which follows from the
basic BCS Hamiltonian (\ref{H}). The latter contains  just one
phenomenological parameter - the Cooper channel interaction
constant, which is encoded in a non-universal ``clean'' transition
temperature, $T_{\rm c0}$. The exact mechanism of unconventional
superconductivity and the exact value of $T_{\rm c0}$ are not
important for the problem at hand, which deals with
disorder-induced suppression of superconductivity that happens at
the diffusive rather than ballistic length-scales. Another
important observation is that action (\ref{S}) is identical to the
Hertz-Moriya-Millis theory~\cite{HMM,Millis} of the
two-dimensional antiferromagnetic QCP. The latter is known to be a
marginal theory~\cite{SK}. Since the dimensionless coupling
constant in Eq.~(\ref{S}) is small, the renormalization group (RG)
treatment of the theory is an asymptotically exact approach.

The upper cut-off in the RG approach is determined by the
applicability of the diffusion approximation, which is reasonable
as long as $|\omega_{\rm max}| \sim D q^2_{\rm max} \lesssim
1/\tau$. In terms of the dimensionless three-momentum $k =
(k_0,{\bf k})$, this condition implies $|k_{\rm max}| = \Lambda
\sim 1$. We use this value of the cut-off and the usual RG
scheme~\cite{Millis}, to obtain the scattering amplitude of
fluctuations $\rho(k_1,k_2,k_3,k_4)$. In principle, the latter
depends on all external momenta, but within the logarithmic
accuracy, we can just consider it to be a function of a single
variable $\rho\left(k = {\rm max}\, \{ k_i \} \right)$, for which
we get the following equation~\cite{SK,Millis} $ \rho(k) = \rho_0
- {5 \over 3} \int_{k} \rho^2(k') \left( r + |{k'}_0| + {{\bf
k}'}^2 \right)^{-2}$, which leads to the ``zero-charge'' behavior
of $\rho(k)$
\begin{equation}
\label{0charge} \rho(k) = {\rho_0 \over 1 + {\displaystyle 5
\rho_0 \over \displaystyle 12 \pi^2} \left|\ln\left[{r + |k_0| +
{\bf k}^2}\right]\right| }.
\end{equation}
Since the bare scattering amplitude is determined by the inverse
conductance, it is alluring to interpret this ``zero charge''
result as a flow of the resistivity to zero in the superconducting
phase. In fact,  a na{\"{\i}}ve calculation of the
Aslamazov-Larkin conductivity diagram~\cite{AL} in which the
current vertices are taken to be independent of frequencies does
lead to the logarithmic behavior of the corresponding correction
to the conductivity~\cite{unpub}, which would be consistent with
the above-mentioned interpretation. However, a more careful
analysis of the Alsamazov-Larkin  diagram shows that this standard
approximation~\cite{LV} (valid near a classical transition) breaks
down near a QCP (see also Ref.~[\onlinecite{GaL1}], where this
complication was first pointed out) and a full frequency and
momentum dependence of the current vertices is needed to get a
correct result. To calculate transport properties turns out to be
very difficult due to the problem of analytical continuation of
the propagators and vertices (which here are very complicated
non-analytic functions of two complex variables). However, to
obtain thermodynamic properties is rather straightforward and can
be done non-perturbatively on the basis of the action (\ref{S}).

Below, we consider the fluctuation magnetic susceptibility near
the quantum phase transition.
\begin{equation}
\label{chi.def} \chi = - {1 \over V} {\partial^2 F \over \partial
H^2} =
 - {1 \over V} \left( {\partial r \over \partial H} \right)^2
 {\partial^2 F \over \partial r^2},
\end{equation}
where the $r(H)$-dependece is given by Eq.~(\ref{QCP}), which
yields $\left( {\partial r / \partial H} \right) = e \tau^2
(v_{\rm F}^2 / c^2)$. Now, we follow Larkin and
Khmelnitskii~\cite{LK} and notice that the second derivative of
the free energy in (\ref{chi.def}) is the exact polarization
operator given by
\begin{equation}
\label{Pi} \Pi(r) = {1 \over D \tau^2} \int\limits {d^3 k \over (2
\pi)^3} {{\cal T}^2(k) \over \left( r + |{k}_0| + {{\bf k}}^2
\right)^2},
\end{equation}
where ${\cal T}(k)$ is the vertex function, which is determined by

\begin{equation}
\label{vertex} {\cal T}(k) = 1 - {2 \over 3} \int\limits_{{\rm
max}\,\{|k_0|,|{\bf k}|\}}^1 {d^3 k' \over (2 \pi)^3} {{\cal
T}(k') \rho(k') \over \left( r + |{k'}_0| + {{\bf k}'}^2
\right)^2}.
\end{equation}
Eqs.~(\ref{chi.def}), (\ref{Pi}), and (\ref{vertex}) lead to the
magnetic susceptibility per unit volume
\begin{equation}
\label{chi.res} \chi = - {12 \pi g^2 \over (3 \overline{u_l^4} -
1)} {e^2 \over d \hbar m c^2} \left\{ \left[ 1 + {5 (3
\overline{u_l^4} - 1) \over 12 \pi^2 g} \ln {1 \over r(T,H,\tau)}
\right]^{1 / 5} - 1 \right\},
\end{equation}
where $d$ is the thickness of the film (or interlayer distance in
the case of a layered superconductor) and the proximity to the
transition $r(T,H,\tau)$ is given by Eq.~(\ref{QCP}).
Eq.~(\ref{chi.res}) has two asymptotic regimes,
\begin{equation}
\label{chi.as} \chi \approx - {e^2 \over d \hbar m c^2}  \left\{
\begin{array}{ll}
{\left( g / \pi \right)}\, \left| \ln {r} \right|,  & \mbox{if }\,  1 \ll |\ln{r}| \ll g;\\
7.358\, g^{9/5} \left| \ln {r} \right|^{1/5}, & \mbox{if }\,
|\ln{r}| \gg g,
\end{array}
\right.
\end{equation}
where the second numerical coefficient corresponds to the $d$-wave
case. The former (finite $r$) behavior is the regime of regular
Gaussian perturbation theory. The later regime of $r \to 0$ is
clearly non-perturbative and becomes asymptotically exact in the
very vicinity of the QCP. Note that  the fluctuation diamagnetism
exceeds the Landau susceptibility $\chi_{\rm Landau} = -{e^2 /( 12
\pi d \hbar m c^2)}$ by orders of magnitude even far from the
transition.
We reiterate that there is no contradiction here~\cite{LV} since
Eq.~(\ref{chi.res}) should be viewed as a correction to the
perfect diamagnetic response of a superconductor (not to the
Pauli/Landau terms in a normal metal). We also note here that the
effect of quantum fluctuations on other thermodynamic properties,
such as the specific heat, are unremarkable since the critical
density of disorder depends on the temperature as $r(T) \propto
T^2$. 

Now we discuss the crossover between the  quantum fluctuation and
 classical fluctuation regimes. Clearly at large temperatures, the
 non-perturbative RG treatment breaks down, because the integral
 over frequency (which makes the quantum problem effectively
 $4$-dimensional) is replaced with the Matsubara sum and only the
 $\omega_n = 0$ term plays a role  near the transition; thus,
 we recover the
 two-dimensional model. The crossover between the two behaviors (here
 we are talking about the linear-response $H \to 0$ magnetic
 susceptibility) happens around $r \sim 2 \pi T \tau = 2\gamma
 (T/T_{\rm co})$. The general expression for the leading order
 correction to the fluctuation susceptibility is determined by
 $\chi^{(1)} = - \left( \partial r / \partial H \right)^2 T
 \sum_{\omega_n} \int d^2{\bf q}/(2\pi)^2 \left[ r + |\omega_n|\tau +
 D{\bf q}^2\tau \right]^{-2}$, which leads to the result
 \begin{equation}
 \label{chi.1}
 \chi^{(1)}(T) = - {e^2 \over d \hbar m c^2} {g \over \pi}
 \left\{  \psi\left[ {1 \over 2 \pi T \tau} \right] -
 \psi\left[ {r(T) \over 2 \pi T \tau }\right] - {\pi T \tau \over r(T)}\right\},
\end{equation}
The low-temperature quantum limit in Eq.~(\ref{chi.1}) reproduces
the logarithmic asymptotic of Eq.~(\ref{chi.as}), while the
``high-temperature'' limit $r \ll (T/T_{\rm c0}) \ll 1$ leads to
the classical Aslamazov-Larkin-type power law divergence
\begin{equation}
\label{chi.1.as}
 \chi^{(1)} = - {3 g\over 4 \pi \gamma} {e^2 \over d \hbar m c^2}
{T_{\rm c0} \over T - T_{\rm c} (\tau)},
\end{equation}
where we introduced a scattering time-dependent transition
temperature [which is the inverse function of $\tau_{\rm c}(T)$
used earlier, see Eq.~(\ref{QCP})]. We emphasize that
Eq.~(\ref{chi.1.as}) corresponds to the limit $(T/T_{\rm c0}) \ll
1$. However, at higher temperatures and in particular near $T_{\rm
c0}$ (clean limit), the familiar Aslamazov-Larkin power law and
all parameters are preserved and only the numerical coefficient
changes.

Fig.~1 summarizes the behavior of fluctuations in the vicinity of
the disorder-induced superconductor-metal transition in an
anisotropic gap superconductor. The solid black line represents
the phase boundary between the superconducting and metallic
phases. The hatched area is the Ginzburg region, where the
fluctuations interact strongly and Aslamazov-Larkin theory breaks
down. Interestingly, the width of the quantum Ginzburg region,
${\rm Gi}_{\rm Q}\, = \exp{(-g)}$, is much smaller than that of
the classical Ginzburg region, ${\rm Gi}_{\rm C}\, = 1/g$. The two
dashed lines in Fig.~1 separate the classical and quantum
fluctuation regimes. The classical regime is effectively a
two-dimensional theory ($d_{\rm eff}^{\rm C} = 2$) in which the
leading order perturbative correction diverges as a power law,
$\chi^{(1)}_{\rm C} \propto -\left(T - T_{\rm c}\right)^{-1}$. The
effective dimensionality in the quantum regime is $d_{\rm
eff}^{\rm Q} = 4$ and the leading order correction to
susceptibility (and likely conductivity~\cite{unpub}) is
logarithmic, $\chi^{(1)}_{\rm Q} \propto \ln\left(n - n_{\rm c}
\right)$. The region in between the two dashed lines represents  a
crossover between the quantum and classical
behavior~\cite{HMM,Millis,QPT.RMP.2} and in some sense describes a
crossover between the effective dimensionality $d_{\rm eff}^{\rm
Q} =4$ in the former to the effective dimensionality $d_{\rm
eff}^{\rm C} =2$ in the latter. The leading order correction to
thermodynamics in this regime is described by the non-linear
function in Eq.~(\ref{chi.1}), which we found to be a smooth
function with no remarkable properties. The exact non-perturbative
behavior in the classical and crossover regions is unknown, but
some insight can be obtained via an
$\epsilon$-expansion~\cite{SK,Millis}.
In the quantum Ginzburg region, the RG approach is asymptotically
exact and leads to the susceptibility, which behaves as $\chi_{\rm
Q} \propto \ln^{1/5}\left(n - n_{\rm c} \right)$.

\begin{figure}
\centering
\includegraphics[width=3.3in]{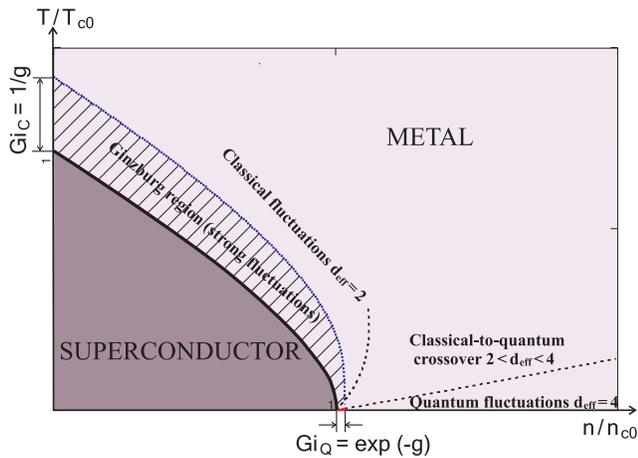}
\caption{\label{FIG:QPT} (Color online) This figure illustrates
the phase diagram (transition temperature {\em vs.} concentration
of disorder) and various fluctuation regimes. The hatched region
is the regime of strong superconducting fluctuations. The dashed
lines determine the quantum, classical, and quantum-to-classical
crossover regions in the fluctuation thermodynamics. See text for
a more detailed discussion.} \vspace*{-0.2in}
\end{figure}

We note that in a  disordered system, the latter non-perturbative
behavior may be smeared out by the Griffiths effects due to
mesoscopic fluctuations of disorder. There always exist
superconducting puddles, which appear in the local areas, where
the concentration of impurities is smaller than the
average~\cite{VG,GaL2,SZ}. At low temperatures, these puddles are
connected via Josephson tunneling,  which is long-range in real
space $J(r_{ij}) \propto |{\bf r}_i - {\bf r}_j|^{-2}$. A global
transition in this random network is governed by phase
fluctuations~\cite{GaL2}.  The corresponding theory is however not
 a pure random $XY$-model. First, in a $d$-wave superconductor (or any other
superconductor where the gap vanishes), there always exist gapless
quasiparticles with a small density of states, which lead to an
unusual intrinsic dissipation and non-local in time Josephson
coupling even in the absence of a normal component (the long-time
asymptote of both the dissipation and Josephson kernels behave as
\cite{dwave.tun} $\alpha(\tau) \propto J(\tau) \propto
|\tau|^{-3}$). For superconducting puddles surrounded by a normal
metal with a large electronic density of states, the dissipation
kernel acquires the usual Caldeira-Leggett form $\alpha(\tau)
\propto |\tau|^{-2}$, while the Josephson term remains long-range
in time $J(\tau) \propto |\tau|^{-3}$. Within a mean field
approach, one can estimate~\cite{unpub} that the QCP~(\ref{QCP})
is shifted by $\delta r \sim g^{-1}$ and therefore the quantum
Ginzburg region gets absorbed by the Griffiths phase. A complete
theory of the quantum Griffiths phase will be presented elsewhere.
We note here that while it is not clear whether the evasive
Griffiths phase can be observed in real experiment, the quantum
fluctuation effects described by the general equation
(\ref{chi.res}) should definitely be accessible at least for
$\delta r \gtrsim g^{-1}$ and possibly in an even wider range of
parameters.

The physical systems where these fluctuations may be
experimentally observed include disordered superconducting films
with an unusual pairing symmetry and possibly  high-$T_{\rm c}$
cuprates in the vicinity of the termination of the superconducting
dome in the overdoped regime. Independently of the nature of the
transition in the overdoped high-$T_{\rm c}$ materials, the
effective theory still should have the form (\ref{S}) [but the
physical meaning of the transition parameter $r$ in (\ref{S}) may
be different] and therefore all functional dependencies should
remain the same. We also note here that recent STM experiments,
(see {\em e.g.} Ref.~\onlinecite{Yazdani}) suggest that disorder
may play an important role in high-$T_{\rm c}$ materials and, if
so, should contribute to $T_{\rm c}$-suppression~\cite{VG}.
Another superconducting system where the field theory of the
transition is identical to (\ref{S}) is an $s$-wave
superconducting film in which the transition temperature is
suppressed to zero by magnetic disorder. Finally, we emphasize
that since a superconductor is a perfect diamagnetic material,
fluctuation diamagnetism  is physically a correction to this
perfect diamagnetic behavior and in the vicinity of a
superconducting transition, the magnetic susceptibility is always
much larger than the Fermi liquid terms. Thus, the experimental
verification of the predicted quantum critical behavior of the
magnetic susceptibility is certainly feasible.

 \vspace*{-0.25in}

\bibliography{QPT}

\end{document}